\documentclass[aps,prl,twocolumn,groupedaddress,longbibliography]{revtex4-1}
\usepackage{graphicx,amsmath,amssymb,color,xcolor}
\begin{document}

\title{Unexpected ductility in semiflexible polymer glasses with entanglement length equal to their Kuhn length}
\author{Joseph D. Dietz}
\author{Kai Nan}
\author{Robert S. Hoy}
\email{rshoy@usf.edu}
\affiliation{Department of Physics, University of South Florida}
\date{\today}
\begin{abstract}
Semiflexible polymer glasses (SPGs), including those formed by the recently synthesized semiflexible conjugated polymers (SCPs), are expected to be brittle because classical formulas for their craze extension ratio $\lambda_{\rm craze}$ and fracture stretch $\lambda_{\rm frac}$ predict that systems with $N_e = C_\infty$ have  $\lambda_{\rm craze} = \lambda_{\rm frac} = 1$ and hence cannot be deformed to large strains.
Using molecular dynamics simulations, we show that in fact such glasses can form stable crazes with $\lambda_{\rm craze} \simeq N_e^{1/4} \simeq C_\infty^{1/4}$, and that they fracture at  $\lambda_{\rm frac} = (3N_e^{1/2} - 2)^{1/2} \simeq (3C_\infty^{1/2} - 2)^{1/2}$.
We argue that the classical formulas for $\lambda_{\rm craze}$ and $\lambda_{\rm frac}$ fail to describe SPGs' mechanical response because they do not account for Kuhn segments' ability to stretch during deformation.
\end{abstract}
\maketitle

One of the most remarkable features of polymer glasses is their ability to form mechanically stable \textit{crazes}, fibrillar loadbearing structures that form ahead of cracks.
Polymer is drawn into a stable craze at a constant stress.
Crazes formed in this way can extend for microns along the direction perpendicular to the crack \cite{kramer83,haward97}.
This allows ductile polymer glasses' fracture energy $G_c$ to be $10^3-10^4$ times larger than their interfacial free energy $G_{\rm eq}$ \cite{brown91}, in contrast to brittle crystalline materials which have $G_c \simeq G_{\rm eq}$.
Their large $G_c$ enables ductile polymer glasses' use in a wide variety of loadbearing applications despite their low elastic moduli compared to metals and other structural materials.

Semiflexible conjugated polymers (SCPs) are attracting great interest owing to their potentially unique combination of electronic and mechanical properties \cite{rivnay13,liao15}.
Many such polymers are semicrystalline and brittle in their solid form, limiting their utility in situations where mechanical ductility and toughness are required \cite{ashizawa20,xie18}.
Other SCPs are better glassformers \cite{xie18,xie20}, but since many of these have only recently been synthesized (and only in small quantities), their melt rheology is only beginning to be studied \cite{fenton22}, and their glassy-state mechanical properties remain largely unknown.
Fortunately, coarse-grained computer simulations can offer key insights into these polymers' potential for loadbearing applications such as wearable and bioimplantable electronic devices \cite{rivnay13,liao15,ashizawa20}.

Key amongst the questions to be answered is whether these polymers are capable of stable craze drawing.
The standard model of craze formation, developed by Kramer \textit{et al.}\ \cite{kramer83,donald82,henkee84}, suggests that they are not.
Kramer's argument proceeds as follows:
The density ratio of undeformed glass and fully-developed crazes (i.e.\ crazes away from the interfacial region) is $\lambda_{\rm craze} = \rho_{\rm u}/\rho_{\rm fd}$.
In the undeformed glass, the mean-squared end-end distance of a typical entangled chain segment is $\langle R^2 \rangle_{\rm u} = \ell_0 \ell_K N_e = \ell_0^2 C_\infty N_e$, where $\ell_0$ is the polymer's backbone bond length, $\ell_K$ is its Kuhn length,  $C_\infty \equiv \ell_K/\ell_0$ is its characteristic ratio, and $N_e$ is its entanglement length.
In a fully developed craze, the segment has pulled taut and its mean-squared end-end distance is $\langle R^2 \rangle_{\rm fd} = \ell_0^2 N_e^2$.
Assuming the segment stretches by a factor $\lambda_{\rm craze}$ as it is drawn into the craze gives   $\langle R^2 \rangle_{\rm fd} = \lambda_{\rm craze}^2 \langle R^2 \rangle_{\rm u}$.

Combining these expressions for  $\langle R^2 \rangle_{\rm fd}$ yields the equation $\ell_0^2 N_e^2 =  \lambda_{\rm craze}^2 \ell_0^2 C_\infty N_e$, which in turn gives the well-known prediction
\begin{equation}
\lambda_{\rm craze} = \sqrt{\displaystyle\frac{N_e}{C_\infty}},
\label{eq:1Dkramer}
\end{equation}
which correctly describes the experimental trends for ductile commodity polymers \cite{kramer83, haward97}.
Remarkably, the $N_e$ that successfully predict $\lambda_{\rm craze}$ are the same as those obtained from measurements of the plateau modulus ($G_N^0 = 4\rho k_B T/5N_e$), indicating that entanglements in polymer glasses are inherited from their parent melts \cite{kramer83}.

An analogous argument can be used to predict the local uniaxial stretch $\lambda_{\rm frac}$ at which the craze must fail via chain scission \cite{kramer83}, leading to brittle fracture of the entire sample.
In the undeformed glass, the mean-squared projections of entangled-segment dimensions along the $x$, $y$, and $z$ axes are
\begin{equation}
\langle R^2 \rangle_{\rm x, u} = \langle R^2 \rangle_{\rm y, u} = \langle R^2 \rangle_{\rm z, u} = \displaystyle\frac{  \ell_0^2 C_\infty N_e}{3}.
\end{equation}
Just before the craze fails, assuming that the sample is uniaxially stretched along the $z$ axis and that entangled segments deform affinely, 
$\langle R^2 \rangle_{\rm x}$ and $\langle R^2 \rangle_{\rm y}$ remain the same, while $\langle R^2 \rangle_{\rm z}$ has increased by a factor of $\lambda_{\rm frac}^2$.
Setting $\langle R^2 \rangle_{\rm fd} = (2 + \lambda_{\rm frac}^2)  \ell_0^2 C_\infty N_e/3 = \ell_0^2 N_e^2$ yields
\begin{equation}
\lambda_{\rm frac} =   \sqrt{\displaystyle\frac{3N_e}{C_\infty} - 2}.
\label{eq:3Dkramer}
\end{equation}
Equation \ref{eq:3Dkramer} accurately predicts $\lambda_{\rm frac}$ in simulations of bead-spring polymer glasses as long as $N_e \gg C_\infty$ \cite{rottler03,nguyen18}.

Equations \ref{eq:1Dkramer} and \ref{eq:3Dkramer} predict $\lambda_{\rm craze} = \lambda_{\rm frac} = 1$ when $N_e = C_\infty$, implying that SPGs, which by definition have $N_e \simeq C_\infty$ \cite{fenton22}, cannot form stable crazes.
Analogous arguments for the natural draw ratio $\lambda_{\rm NDR}$ predict $\lambda_{\rm NDR}^2 + 2/\lambda_{\rm NDR} = 3N_e/C_\infty$ and hence $\lambda_{\rm NDR} = 1$,  implying that they cannot form stable shear bands either.
If SPGs cannot yield via either crazing or localized shear banding, they are very likely to be brittle, severely limiting their utility for load-bearing applications.

In this Letter, using molecular dynamics simulations of a bead-spring model which has been shown to capture many features of glassy-polymeric mechanical response \cite{rottler09,roth16}, we show that in fact SPGs with $N_e \simeq C_\infty$ \textit{can} form stable crazes, with $\lambda_{\rm craze} \simeq N_e^{1/4} \simeq C_\infty^{1/4}$, and that they fracture at  $\lambda_{\rm frac} \simeq (3N_e^{1/2} - 2)^{1/2} \simeq (3C_\infty^{1/2} - 2)^{1/2}$, over a wide range of temperatures.
For low $T$, the microstructure of these crazes is quantitatively but not qualitatively different from those formed in flexible polymer glasses (FPGs).
For higher $T$, cavitation is more localized -- voids grow in a non-system-spanning fashion -- but the coexistence of unyielded and cavitated regions 
at equal stress stabilizes systems against fracture in a manner similar to that recently observed in experiments on very-ductile, densely entangled FPGs \cite{charvet19,djukic20}.
We explain these results by (i) recognizing that  Kuhn segments in undeformed glasses are not straight, but in fact consist of $\sqrt{C_\infty}$ statistical segments and hence can pull taut during craze extension, and (ii) arguing that SPGs can be mechanically stabilized by the combination of a high rate of void nucleation (cavitation) combined with a relatively low rate of void growth.

We study crazing in model SPGs using molecular dynamics (MD) simulations of the semiflexible, breakable-bond variant of the Kremer-Grest model \cite{kremer90,faller99,stevens01}.
All MD simulations are performed using LAMMPS \cite{thompson22}.
Monomers have mass $m$ and interact via the truncated and shifted Lennard-Jones potential $U_\textrm{LJ}(r) = 4\varepsilon[(a/r)^{12} - (a/r)^{6} - (a/r_{c})^{12} + (a/r_c)^{6}]$, where $\varepsilon$ is the intermonomer binding energy, $a$ is the monomer diameter, and $r_c = 2^{7/6}a$ is the cutoff radius. 
The Lennard-Jones time unit is $\tau = \sqrt{ma^2/\varepsilon}$, and the MD timestep employed in this study is $\delta t = \tau/200$.
Covalent bonds are modeled using a quartic potential commonly employed in studies of glassy-polymeric fracture \cite{stevens01,rottler09,nguyen18,footB2}:
\begin{equation}
U_{\rm bond}(\ell) =  k_q(\ell-R_b)^3 (\ell-R_b-B_2).
\label{eq:Uqua}
\end{equation}
Bonds break when their length ($\ell$) exceeds $R_b = 1.3a$.
The ratio of the forces at which covalent and van der Waals bonds break is set to $50$ by setting $k_q = 4431\varepsilon/a^4$; this choice makes bond scission slightly easier than in many previous studies \cite{rottler03,nguyen18}.
Angular interactions between three consecutive beads along chain backbones are modeled using the standard potential $U_{\rm ang}(\theta) = \kappa[1 - cos(\theta)]$ \cite{faller99}, where $\theta$ is the angle between consecutive bond vectors.
Here we primarily employ well-entangled $\kappa = 5.5\varepsilon$ chains, which (at temperature $T = \varepsilon/k_B$) have $C_\infty \simeq 10.3$ and $N_e \simeq 10.3$ \cite{dietz22}.
This value of $\kappa/k_B T$ places systems just below the onset of mid-range nematic order, where chains are maximally entangled \cite{faller99,hoy20,dietz22b}.
The same stiffness regime is occupied by some of the SCP melts studied in Refs.\ \cite{xie18,xie20,fenton22,xie18b}.

Polymer melts composed of $N_{ch}=1000$ linear chains of $N = 400$ monomers were thoroughly equilibrated at $T = \varepsilon/k_B$ as described in Ref.\ \cite{dietz22}, and then slowly cooled at zero pressure as described in Ref.\ \cite{nguyen18}.
After cooling, we uniaxially extended systems along their $z$-axes at a constant true strain rate $\dot\epsilon \equiv \dot\epsilon_{zz} = 10^{-5}/\tau$ that is small enough to be near the quasistatic limit \cite{footLang}.
We deformed systems at this rate until they had extended well beyond fracture, as identified by the postyield maximum in the axial stress $\sigma_{zz}$.
Postyield cavitation and void growth during these runs were characterized using  a refined version \cite{voidid} of the method described in Ref.\ \cite{nan21}.

\begin{figure}[h!]
\includegraphics[width=3in]{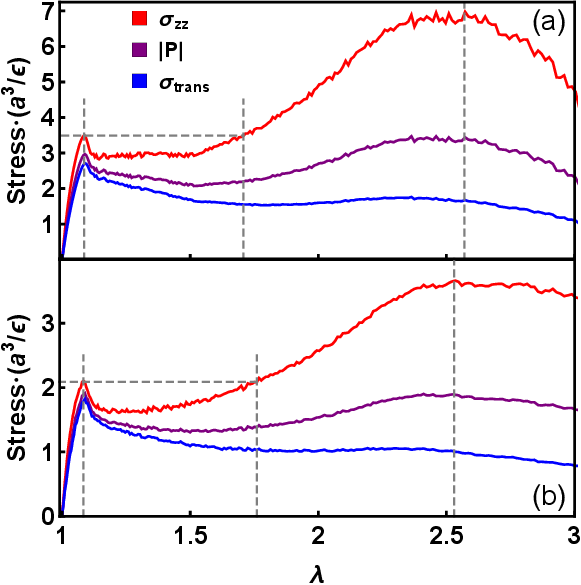}
\caption{ Stress-strain curves for our model SPG at $T = 0$ [panel (a)] and $T \simeq 3T_g/4$ [panel (b)]. 
Vertical dashed lines from left to right respectively correspond to $\lambda = \lambda_{\rm yield}$, $\lambda = \lambda_{\rm craze}$ and $\lambda = \lambda_{\rm frac}$, and the horizontal dashed lines illustrate the coexistence of unyielded and crazed regions at equal $\sigma_{\rm zz}$.}
\label{fig:stressstrain}
\end{figure}

\begin{figure}[htbp]
\includegraphics[width=2.95in]{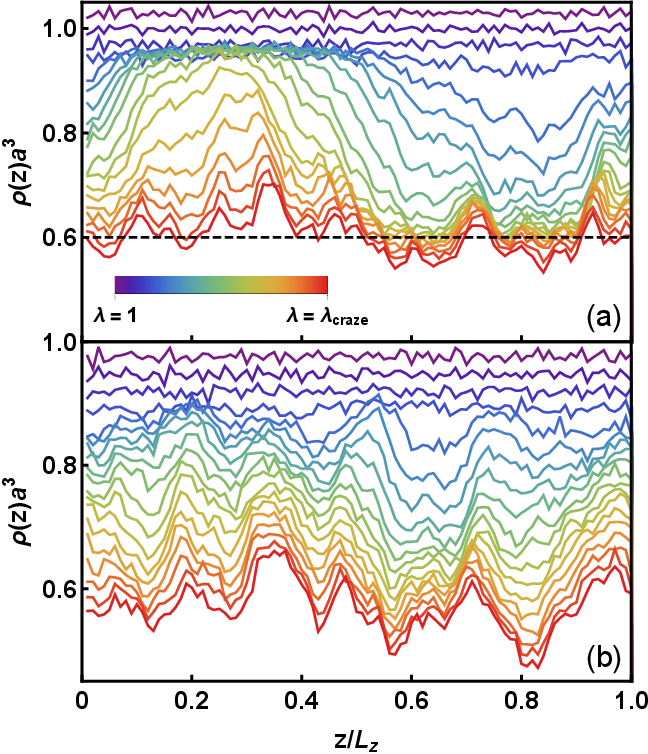}
\caption{Density profiles vs. strain for our model SPG at $T = 0$ [panel (a)] and $T \simeq 3T_g/4$ [panel (b)].  The dashed black line in panel (a) indicates $\rho_{\rm fd}$. Here the $L_z = \lambda L_z^0$ are the lengths of the periodic simulation cells along their $z$-axes.}
\label{fig:densprofiles}
\end{figure}

First we discuss the basic features of this model SPG's mechanical and structural responses to deformation at two temperatures [$T = 0$ and $T = 0.44\varepsilon/k_B \simeq 3T_g/4$] that should respectively favor brittle and ductile deformation \cite{berger88,roth16}; the latter corresponds to $T$ above $T_{room}$ for many SCPs \cite{xie18,xie20,fenton22}.
We also performed deformation simulations at other $0 < T < 3T_g/4$, and found that all results were intermediate between those discussed below.
Figure \ref{fig:stressstrain} shows that the axial stress $\sigma_{\rm zz}(\lambda)$, transverse stress $\sigma_{\rm trans}(\lambda) = [\sigma_{\rm xx}(\lambda) + \sigma_{\rm yy}(\lambda)]/2$, and pressure $P =  [\sigma_{\rm xx}(\lambda) + \sigma_{\rm yy}(\lambda) + \sigma_{\rm zz}(\lambda)]/3$, where $\lambda \equiv L_z/L_z^0 \equiv \ln(\epsilon_{zz})$ is the uniaxial stretch, are all qualitatively identical to those exhibited by flexible polymer glasses undergoing stable crazing in both experiments and simulations \cite{kramer83,haward97,rottler03,rottler09}, for both $T$.
Specifically, $\sigma_{\rm zz}$  exhibits an elastic regime, followed by very sharp cavitation-induced yielding and massive strain softening at $\lambda = \lambda_{\rm yield}$, followed by stable craze drawing at nearly constant stress, followed by strain hardening that is roughly linear in $\lambda$, and finally by fracture at $\lambda = \lambda_{\rm frac}$, with $\lambda_{\rm frac} = 2.57$ for $T = 0$ and $2.53$ for $T = 3T_g/4$.
For mechanically stable crazes, one expects $\lambda_{\rm craze}$ to satisfy $\sigma_{\rm zz}(\lambda_{\rm yield}) = \sigma_{\rm zz}(\lambda_{\rm craze})$; the horizontal dash-dotted lines suggest $\lambda_{\rm craze} \simeq 1.71$ for $T = 0$ and $1.76$ for $T = 3T_g/4$.

Next we show explicitly that this model SPG can form stable crazes, as indicated by the presence of coexisting high- and low-density regions with density ratio $\sim \lambda_{\rm craze}$.
Figure \ref{fig:densprofiles} shows how the monomer-number-density profiles $\rho(z/L_z)$ evolve with increasing strain.
In the undeformed glass, $\rho = \rho_{\rm u}$ is $z$-independent.
At $\lambda = \lambda_{\rm yield} \simeq 1.09$, as yielding occurs via cavitation and craze nucleation \cite{rottler03}, one or more lower-density regions forms.
For the $T = 0$ glass [panel (a)], a single low-density region forms, centered at $z \simeq 0.8L_z$.
The density of this region continues to decrease as more material is drawn into the craze, until it plateaus at $\rho = \rho_{\rm fd} \simeq 0.60a^{-3}$.
Then it remains nearly constant as  the rest of the polymer is progressively drawn into the craze.
Thus this glass has $\lambda_{\rm craze} \simeq 1.03/0.6 \simeq 1.72$.
For the $T = 3T_g/4$ glass [panel (b)], the above picture is obscured because cavitation and void growth is less localized along the axial direction.
However, the similarity of the stress-strain curves shown in Fig.\ \ref{fig:stressstrain} suggests that essentially the same physics is controlling both systems.
In particular, they suggest that upon yielding via cavitation, localized regions within samples stretch by a factor $\sim \lambda_{\rm craze}/\lambda_{\rm yield}$ before being stabilized by these systems' strong strain hardening.

\begin{figure}[h]
\includegraphics[width=3in]{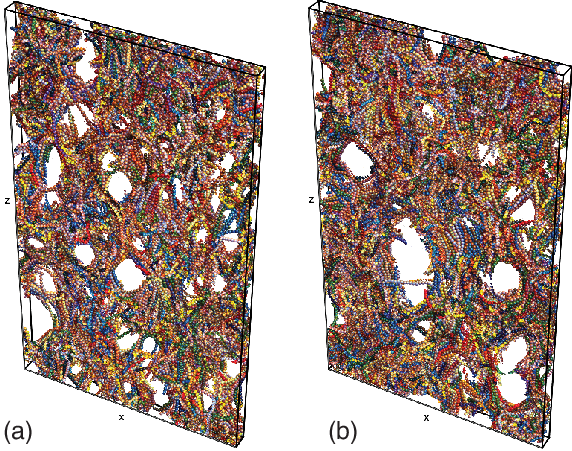}
\caption{ Cross-sections of thickness $5\sigma$ illustrate the structure of our SPG at $\lambda = \lambda_{\rm craze}$, for $T = 0$ [panel (a)] and $T = 3T_g/4$ [panel (b)].}
\label{fig:structpics}
\end{figure}

Crazes in flexible-polymer glasses (FPGs), all of which have $N_e \gtrsim 4 C_\infty$ \cite{kramer83,mark07}, have a universal microstructure.
Their primary fibrils, which are highly oriented along the direction of extension, have a typical diameter $D$ and are separated by a typical distance $D_0$.
These fibrils are held together by entanglements, and are further stabilized by smaller cross-tie fibrils \cite{kramer83,haward97}.
Figure \ref{fig:structpics} shows that for low $T$, the tighter entanglement constraints present in SPGs modify this structure quantitatively but not qualitatively.
Compared to FPG crazes \cite{rottler03,rottler09,ge17}, SPG crazes have a larger $D/D_0$, fibrils that are less oriented along the $z$-direction, and less-clear distinctions between primary and cross-tie fibrils.
For higher $T$, the dilatation inherent to the uniaxial extension employed in this study occurs through nucleation and growth of intermediate-sized voids that do not span the $xy$ plane, rather than through drawing of a single craze.
Such voids are consistent with a combination of cavitation and shear yielding, i.e.\ with the strain localization mechanism expected for a densely entangled ductile FPG \cite{kramer83,haward97}.

\begin{figure}[h]
\includegraphics[width=3in]{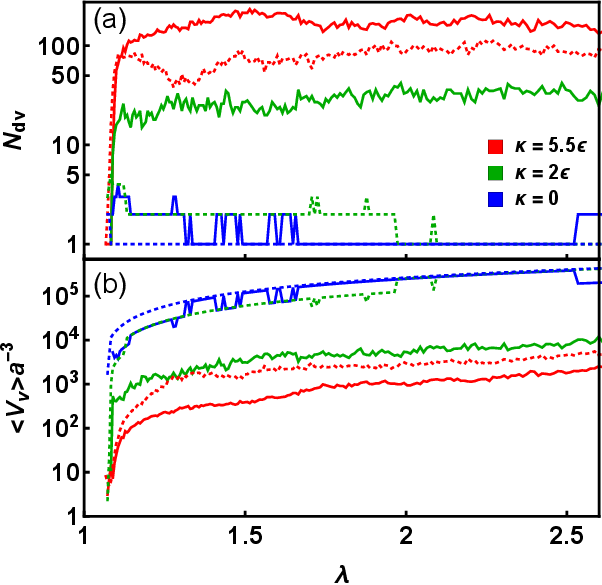}
\caption{ Number of topologically distinct voids $N_{\rm dv}$ [panel (a)] and average void volumes $\langle V_{\rm v} \rangle$ [panel (b)] in model SPGs and in model FPGs with $\kappa = 0$ and $\kappa = 2\varepsilon$ \cite{nan21}.  
Solid and dotted curves respectively show results for $T = 0$ and $T = 3T_g/4$.}
\label{fig:ndv}
\end{figure}

Recent experiments on extremely-ductile polymer glasses like polyphthalamide (PPA) \cite{charvet19,djukic20} have shown that formation of voids ahead of craze fronts effectively dissipates energy and promotes ductility.
The same experiments have shown that void growth that is linear or sublinear in $\lambda$ indicates mechanical stabilization by strain hardening (which has long been known to promote ductility \cite{kramer83,haward97}), in contrast with unstable supralinear void growth that leads to fracture.
Figure \ref{fig:ndv} shows that both these trends are displayed by our model SPGs.
After cavitation occurs at $\lambda \simeq \lambda_{\rm yield}$, systems' effective void volume fractions $f_v \sim (\lambda - 1)/\lambda$ do not depend strongly on chain stiffness, but SPGs form far more (and hence far smaller) voids than FPGs for both $T$.
Many of these voids nucleate ahead of the craze front or larger non-system-spanning voids, and remain very small.
Growth of the largest voids appears to be linear in $\lambda$ for $\lambda_{\rm yield} < \lambda \lesssim \lambda_{\rm frac} - 0.2$.
Fewer (and larger) voids form for $T = 3T_g/4$ than for $T = 0$, consistent with the contrast illustrated in Fig.\ \ref{fig:structpics}.
Overall these results suggest that the same micromechanisms that stabilize PPA \cite{charvet19,djukic20} can also stabilize suitably designed SPGs.

The surprising ability of our SPGs to stably craze-draw can be explained by recognizing that, contrary to the assumptions underlying Eqs.\ \ref{eq:1Dkramer} and \ref{eq:3Dkramer},  $N_e \simeq C_\infty$ does \textit{not} imply that entangled strands cannot substantially stretch during sample deformation.
A typical Kuhn segment contains $\sqrt{C_\infty}$ statistical segments of length $b = \sqrt{\ell_0 \ell_K} = \sqrt{C_\infty} \ell_0$ \cite{bdef}.
In the undeformed glass, its mean-squared end-end distance is $\langle R^2 \rangle_{\rm u} = b^2 \sqrt{C_\infty} = \ell_0^2 C_\infty^{3/2}$.
Pulling this segment taut increases its mean-squared end-end distance to $\ell_0^2 C_\infty^2$.
Assuming that this taut state corresponds to fully developed crazing gives us $\langle R^2 \rangle_{\rm fd} = \lambda_{\rm craze}^2 \langle R^2 \rangle_{\rm u}$, which yields $\lambda_{\rm craze} = C_\infty^{1/4}$.
For SPGs with $N_e = C_\infty$, this expression becomes \cite{expNe}
\begin{equation}
\lambda_{\rm craze} = N_e^{1/4} \equiv \left( \displaystyle\frac{N_e }{\sqrt{C_\infty}}   \right)^{1/2}.
\label{eq:lamsf2}
\end{equation}
Following the same arguments that were used to obtain Eq.\ \ref{eq:3Dkramer} yields an analogous formula for $\lambda_{\rm frac}$:
\begin{equation}
\lambda_{\rm frac} =  \left( 3 \displaystyle\frac{N_e }{\sqrt{C_\infty}} - 2  \right)^{1/2} \equiv  \left( 3 C_\infty^{1/2}  - 2  \right)^{1/2}.
\label{eq:lamfrac2}
\end{equation}
For the $\kappa = 5.5\epsilon$, $N_e = C_\infty = 10.3$ chains considered here, Eqs.\ \ref{eq:lamsf2}-\ref{eq:lamfrac2} predict $\lambda_{\rm craze} = 1.79$ and $\lambda_{\rm frac} = 2.75$.
The former prediction  is quantitatively consistent with our measured $\lambda_{\rm craze} \simeq 1.7-1.8$.
While the latter is slightly above our measured  $\lambda_{\rm frac} = 2.5-2.6$, a difference of this magnitude is expected because chains break (via scission) before their Kuhn segments pull completely taut.

Strictly speaking, Eqs.\ \ref{eq:lamsf2}-\ref{eq:lamfrac2} should accurately predict $\lambda_{\rm craze}$ and $\lambda_{\rm frac}$ only when $N_e = C_\infty$.
One expects a gradual crossover to validity of the Kramer formulas with increasing $N_e/C_\infty$, satisfying \cite{prelimk5}
\begin{equation}
\sqrt{\displaystyle\frac{N_e}{C_\infty}} \lesssim \lambda_{\rm craze}  \lesssim \left( \displaystyle\frac{N_e }{\sqrt{C_\infty}}   \right)^{1/2}.
\label{eq:lamcg}
\end{equation}
Future work will aim to identify both a universal expression for $\lambda_{\rm craze}$ and the micromechanisms underlying its functional form.
One potential explanation for the crossover proceeds as follows: 
The total number of energetically-costly dihedral rearrangements required to affinely stretch all the chain segments (of chemical length $n$) in a polymer glass by a factor $\Lambda$ scales roughly as $\Lambda n^{-1/2}$.
Thus the work required to stretch a macroscopic sample by a factor $\lambda$ is minimized when chain stretching is very subaffine  (i.e.\ has $\Lambda \ll \lambda$) at small $n$ and affine (i.e.\ has $\Lambda = \lambda$) only for $n \gtrsim N_e$. 
This implies that Kuhn segments in a fully developed craze should be very weakly (strongly) stretched
in systems with $N_e \gg C_\infty$ ($N_e \simeq C_\infty$), and could explain why Eqs.\ \ref{eq:1Dkramer} and \ref{eq:3Dkramer} (Eqs.\ \ref{eq:lamsf2} and \ref{eq:lamfrac2}) accurately predict FPGs' (SPGs') $\lambda_{\rm craze}$ and $\lambda_{\rm frac}$.

Designing highly ductile SCPs that are well-suited for electronics applications has been challenging because high conductivity (high ductility) usually requires a high (low) degree of crystallinity \cite{xie18,ashizawa20}.
Factors such as the prevalence of intermolecular $\pi-\pi$ stacking, the degree of regioregularity, and side chain architecture also play important roles. 
For these reasons, much recent work has focused on designing polymers that include the structural features promoting high conductivity, yet are primarily amorphous \cite{ashizawa20,xie18,son18}.
The bead-spring model employed in this study does not crystallize, and does not account for the factors that control conductivity.
Nonetheless we claim that the above results provide useful guidance for SCP/SPG design, as follows:

Increasing SCPs' $C_\infty$ tends to improve their conductivity \cite{xie18}.
Decreasing FPGs' $N_e$ leads to larger strain hardening moduli $G_{\rm R}$, which stabilize systems against brittle fracture \cite{kramer83,haward97}.
This combination of factors suggests maximizing $C_\infty$ while minimizing $N_e$, but increasing $C_\infty$ too much leads to nematic ordering that increases $N_e$ \cite{xie18b,fenton22}.
Since the highest $G_N^0$ are achieved in SCPs with  $N_e \lesssim 2C_\infty$ \cite{fenton22}, and $G_{\rm R} \propto G_N^0$ \cite{vanmelick03,kramer05}, it seems reasonable to suggest that SPGs with $N_e \simeq C_\infty$, formed, for example, by SCPs that are more densely entangled than the widely employed P3HT \cite{ashizawa20,xie18,xie20,fenton22}, should have optimal mechanical performance.
On the other hand, Kramer's classic picture \cite{kramer83}, which is widely accepted as correct for FPGs \cite{haward97,roth16}, predicts that such systems will be brittle because they have $\lambda_{\rm craze} = \lambda_{\rm frac} = 1$.

Here we have shown that this need not be the case.
Model SPGs with $N_e = C_\infty$ can stably craze-draw, and exhibit a mechanical response that is qualitatively the same as that of their flexible counterparts.
They appear to be stabilized against fracture by their strong strain hardening and by void-formation mechanisms similar to those observed in very-ductile FPGs \cite{charvet19,djukic20}.
We have explained this result in terms of the Kramer theory's failure to account for chain stretching at the Kuhn-segment scale, and developed alternative theoretical expressions for  $\lambda_{\rm craze}$ and  $\lambda_{\rm frac}$ [Eqs.\ \ref{eq:lamsf2}-\ref{eq:lamfrac2}] that quantitatively match our simulation results. 
The extensive track record of bead-spring models in successfully explaining previously-poorly-understood aspects of glassy polymer mechanics \cite{rottler09,roth16} suggests that these expressions should predict the response of the amorphous regions within solid SCPs.

We emphasize, however, that brittle fracture in MD simulations of glassy polymer mechanics can be artificially suppressed by their use of periodic boundary conditions (and consequent lack of the surface defects at which fracture of real systems usually initiates \cite{sauer83}) as well as the small system sizes \cite{zhang20} and fast thermal-quench rates \cite{ozawa18} to which currently-available computational power restricts them.
Such effects may make the ductile response reported above challenging to observe in practice, making it especially important for experimentalists interested in developing new ductile SPGs to employ small sample sizes, rapid thermal quenching to temperatures that are only slightly below $T_g$, and sample-preparation techniques that minimize surface flaws \cite{li15}.

We thank Ralph H.\ Colby for helpful discussions.
This material is based upon work supported by the National Science Foundation under Grant No.\ DMR-1555242.


\begin{thebibliography}{39}
\bibitem [{foo({\natexlab{a}})}]{footB2}%
  \BibitemOpen
  \href@noop {} {}\bibinfo {note} {$B_2$ is selected by
  matching $U_{qu}(\ell)$ to the first zero and the minimum of the commonly
  employed FENE potential \cite{kremer90}, which sets $B_2=
  -0.4668a$.}\BibitemShut {Stop}%
\bibitem [{foo({\natexlab{b}})}]{footLang}%
  \BibitemOpen
  \href@noop {} {}\bibinfo {note} {Temperature was held
  nearly constant during all deformation runs, using a Langevin thermostat with
  damping time $\tau_{\rm Lang} = 10\tau$. This kept the kinetic temperature
  below $10^{-3}\varepsilon/k_B$ during the $T_{\rm target} = 0$
  runs.}\BibitemShut {Stop}%
\bibitem [{voi()}]{voidid}%
  \BibitemOpen
  \href@noop {} {}\bibinfo {note} {We defined a void as a volume containing at
  least two adjacent cubic subcells of volume $\sim a^3$ that contain no
  monomer cores; see Ref.\ \cite{nan21}.}\BibitemShut {Stop}%
\bibitem [{bde()}]{bdef}%
  \BibitemOpen
  \href@noop {} {}\bibinfo {note} {Here $b$ is defined through the classic relation $\langle R_{ee}^2 \rangle = Nb^2$, where $R_{ee}$ is a chain's end-end distance; $b$ can be directly measured in scattering experiments \cite{mark07}.}\BibitemShut {Stop}%
\bibitem [{exp()}]{expNe}%
  \BibitemOpen
  \href@noop {} {}\bibinfo {note} {Note that while experimental studies often define $N_e$ as the number of statistical segments per entangled strand \cite{mark07} rather than the number of monomers per entangled strand as we have done here, this alternative definition does not affect the above arguments; it reduces the $N_e$ in Eqs.\ \ref{eq:lamsf2}-\ref{eq:lamcg} by a factor of $\sqrt{C_\infty}$ and then sets $C_\infty = 1$.}\BibitemShut {Stop}%
\bibitem [{pre()}]{prelimk5}%
  \BibitemOpen
  \href@noop {} {}\bibinfo {note} {Preliminary results for SPGs with $N_e/C_\infty > 1$ are consistent with this hypothesis.  For example, SPGs with $\kappa = 5\varepsilon$, which have $N_e \simeq 10.4$ and $C_\infty \simeq 9.3$  \cite{dietz22b}, have $\lambda_{\rm craze} = 1.83$, which lies between $\sqrt{N_e/C_\infty} = 1.06$ and $(N_e/\sqrt{C_\infty})^{1/2} = 1.85$.}\BibitemShut {Stop}%
\end{thebibliography}

%

\end{document}